\documentstyle[preprint,aps]{revtex}

\begin{document}
\draft

\title{$\eta d$ scattering in the region of the $S_{11}$ resonance} 

\author{H. Garcilazo $^{(1),(3)}$
and M. T. Pe\~na $^{(2),(3)}$}

\address{$(1)$ Escuela Superior de F\' \i sica y Matem\'aticas \\
Instituto Polit\'ecnico Nacional, Edificio 9,
07738 M\'exico D.F., Mexico}

\address{$(2)$ Centro de F\'\i sica das Interac\c c\~oes Fundamentais, \\
Av. Rovisco Pais, P-1049-001 Lisboa, Portugal }

\address{$(3)$ Departamento de F\'\i sica, Instituto Superior T\'ecnico,\\
Av. Rovisco Pais, P-1049-001 Lisboa, Portugal}

\maketitle

\begin{abstract}
We have studied the reaction $\eta d \to \eta d$
close to threshold within a nonrelativistic three-body formalism. 
We considered several $\eta$N and NN models, in particular potentials with 
separable form, fitted to the low-energy $\eta N$ and $NN$ data 
to represent the two-body interactions. We found that with realistic
two-body interactions a
quasibound state does not exist in this system, although there is an
enhancement of the cross section by one order of magnitude,
in the region near threshold, which is a genuine three-body effect not
predicted within the impulse approximation.

\end{abstract}

\pacs{14.40.Ag,25.40.Ve,25.80.Hp}

\narrowtext
\newpage

\section{Introduction}

The elastic scattering $\eta d$ reaction has been studied recently by several
authors \cite{UEDA,RAKI,GRE1,SHEV}
in order to investigate the existence of a resonance or a quasibound
state in this system, for which there are experimental
indications
\cite{PLOU,CAL1,CALE,METAG}. Some of these studies concluded that 
such an state would exist for certain values of the two-body $\eta$N data.
However, since they used in one form or another incomplete
information on the two-body subsystems (in particular that corresponding
to the $\eta N$ sector), we believe that a new calculation is required
which takes into account all the information that is now available. In 
particular, we study the effect of the repulsion at short distances of
the $NN$ interaction and take into account the
$\eta N \to \eta N$ scattering amplitude that has been
determined recently \cite{BAT1,BAT2,GRE2,GRE3}. 

We will present the Faddeev formalism in section II. In section III we  
will first calculate the eta-deuteron scattering length to compare  with 
the results of multiple-scattering theories as well as with
separable potentials models and finally we will
present the predictions of our model for $\eta d$ scattering. We
will give our conclusions in section IV.

\section{Faddeev Formalism}

Let us consider a  system of three particles, where
two of them are identical, interacting
pairwise through separable potentials that act only in S-waves. In
the case of the $\eta d$ system, S-wave means, for the eta-nucleon pair the
$S_{11}$ channel, and for the nucleon-nucleon pair the $^3S_1$ channel. The
two-body T-matrix of the pair $jk$ will be assumed of the separable form

\begin{equation}
t_i(p_i,p^\prime_i;E)=g_i(p_i)\tau_i(E)g_i(p^\prime_i),
\label{eq1}
\end{equation}
where $\tau_i(E)$ and $g_i(p_i)$ will be specified later.
In the following 
we will identify particle 1
with the $\eta$ and the identical particles 2 and 3 with the two nucleons.

The Faddeev equations for the case of $\eta d$ scattering can be solved
explicitely for the $\eta d \to N (N \eta)$ transition  amplitude $T_2$, with
nucleon 2 being the spectator particle in the final state. One obtains

\begin{equation}
T_2(q_2;E) = 2K_{21}(q_2,q_{10};E) + \int_0^\infty {q_2^\prime}^2 dq_2^\prime
M(q_2,q_2^\prime;E)\tau_2(E-{q_2^\prime}^2/2\nu_2)T_2(q_2^\prime;E),
\label{eq2}
\end{equation}
where 

\begin{equation}
M(q_2,q_2^\prime;E) = K_{23}(q_2,q_2^\prime;E) + 
2\int_0^\infty q_1^2 dq_1
K_{21}(q_2,q_1;E)\tau_1(E-q_1^2/2\nu_1)K_{12}(q_1,q_2^\prime;E),
\label{eq3}
\end{equation}
and the driving term for the transition from the amplitude with spectator j
in the initial state to the one with spectator i in the final state  is 

\begin{equation}
K_{ij}(q_i,q_j;E) =  
{1\over 2}\int_{-1}^1 dcos\theta {g_i(p_i)g_j(p_j)\over
E-p_i^2/2\mu_i-q_i^2/2\nu_i+i\epsilon},
\label{eq4}
\end{equation}
with

\begin{equation}
p_i=\left({\mu_i^2\over m_k^2}q_i^2+q_j^2+2{\mu_i\over m_k}q_i q_j 
cos\theta\right)^{1/2},
\label{eq5}
\end{equation}

\begin{equation}
p_j=\left({\mu_j^2\over m_k^2}q_j^2+q_i^2+2{\mu_j\over m_k}q_i q_j 
cos\theta\right)^{1/2}.
\label{eq6}
\end{equation}
$\mu_i$ and $\nu_i$ are the reduced masses

\begin{equation}
\mu_i = {m_j m_k \over m_j + m_k},
\label{eq7}
\end{equation}

\begin{equation}
\nu_i = {m_i(m_j+ m_k) \over m_i+ m_j + m_k},
\label{eq8}
\end{equation}
and the $\eta d$ on-shell relative momentum $q_{10}$ is defined by the
relation

\begin{equation}
E={q_{10}^2 \over 2\nu_1} - B_d,
\label{eq9}
\end{equation}
where $B_d$ is the binding energy of the deuteron.

The $\eta d$ elastic-scattering amplitude is obtained from the solution
of Eq. (\ref{eq2}) by inserting into it a final $\eta d$ state, as 

\begin{equation}
F_{\eta d}(q_{10})=-{\pi\nu_1 \over N}\int_0^\infty q_2^2 dq_2
K_{12}(q_{10},q_2;E)\tau_2(E-q_2^2/2\nu_2)T_2(q_2;E).
\label{eq10}
\end{equation}
where $N$ is the normalization of the deuteron wave function

\begin{equation}
N=\int_0^\infty p_1^2 dp_1 \left[g_1(p_1)\over B_d+p_1^2/2\mu_1\right]^2.
\label{eq11}
\end{equation}
The $\eta d$ scattering length is given by

\begin{equation}
A_{\eta d}=F_{\eta d}(0),
\label{eq12}
\end{equation}
while the integrated elastic cross section is given by

\begin{equation}
\sigma_{ELAS} =4\pi|F_{\eta d}(q_{10})|^2.
\label{eq14}
\end{equation}

Notice that Eqs. (\ref{eq2}) - (\ref{eq4}) and (\ref{eq10})
do not include
the $\pi N$ channel explicitely, but only 
through the inelasticity of the $\eta N$ channel. As for the
the $\eta N$ inelasticity due to the $\pi\pi N$ channel,
its contribution  is not yet 
included at this stage of the calculations.

\section{Results}

We started by calculating
the $\eta d$ scattering length (\ref{eq12}), by solving the
integral equation (\ref{eq2}) with the method of matrix inversion after 
replacing the integration with
a gaussian  mesh. Notice that in this case Eqs.
(\ref{eq2})-(\ref{eq4})
are free of singularities ( the singularity of ${\tau}_1$ at $E=-B_d$
in eq. (\ref{eq3}) occurs for $q_1=0$ and is regularized by the integration volume element). 
Additionally, we also calculated the integrated elastic cross section of $\eta d$
scattering. In order to solve the integral equation (\ref{eq2})  above
threshold, we used the method of contour rotation \cite{HETH}. 

\subsection{$A_{\eta d}$ with non-dynamical separable models}

We will calculate here the $\eta d$ scattering length $A_{\eta d}$ 
for the models proposed in Refs. \cite{GRE1,SHEV}. The signal that a
quasibound state exists for a given model is that the real part of
$A_{\eta d}$ becomes negative while the imaginary part gets large.

Notice that in Eq. (\ref{eq1}) we have assumed a separable model
for the two-body amplitudes $t_i$. This form of the T-matrix is
obtained if one assumes a separable potential between particles 
$j$ and $k$: a dynamical two-body
equation determines the function $\tau_i(E)$
by the form factors $g_i(p_i)$, which carry information on the the range
of the potential, and the strength 
parameter of the potential. However, in the AGS formalism \cite{AGS} used
in Ref. \cite{SHEV} and in the multiple-scattering approach used
in Ref. \cite{GRE1} the function $\tau_2(E)$ for the $\eta N$
subsystem, instead of being calculated, has been chosen
independently of the form factor $g_2(p_2)$.
Such an assumption violates the spirit of the Faddeev approach which
requires a two-body interaction in order to relate the off-shell
behavior of the T-matrix in the energy variable $E$ to the off-shell
behavior in the momentum variable $p_2$. Nevertheless, it is 
instructive to repeat those calculations in order to check the 
accuracy of our numerical solution by comparing with the exact 
result of Ref. \cite{SHEV} as well as to test the convergence of the
multiple-scattering schemme developed in \cite{GRE1} which is based
in a partial summation of the multiple-scattering series. 

In both Refs. \cite{GRE1,SHEV} the form factor $g_2(p_2)$ has been
taken of the Yamaguchi form

\begin{equation}
g_2(p_2)={1 \over \alpha_2^2 + p_2^2},
\label{eq17}
\end{equation}
with $\alpha_2=3.316$ fm$^-1$. In Ref. \cite{SHEV}, the function 
$\tau_2(E)$ has been parametrized as

\begin{equation}
\tau_2(E) = {\lambda_\eta \over E-E_0 + {i\over 2}\Gamma},
\label{eq18}
\end{equation}
with $E_0=1535$ MeV - $(m_N+m_\eta)$ and $\Gamma=150$ MeV. The 
parameter $\lambda_\eta$ in Eq. (\ref{eq18}) was chosen to reproduce the
complex $\eta N$ scattering length $a_{\eta N}$, by using the relation

\begin{equation}
t_2(0,0;0)=-{a_{\eta N} \over \pi \mu_2},
\label{eq19}
\end{equation}
where $\mu_2$ is the $\eta N$ reduced mass. This leads to 

\begin{equation}
\lambda_\eta={\alpha_2^4(E_0-i\Gamma/2) \over \pi \mu_2}a_{\eta N}.
\label{eq20}
\end{equation}
As for the nucleon-nucleon separable T-matrix used in Ref. \cite{SHEV},
it was generated from a Yamaguchi separable potential with an
energy-dependent strength \cite{SHEV,GAR1}. Using these parameters
we calculated the $\eta d$ scattering length $A_{\eta d}$ with the
formalism described in the previous section for a variety of values
of the $\eta N$ scattering length $a_{\eta N}$ that have been proposed
in the literature (see Refs. \cite{GRE1,SHEV}). 
We show the results of this comparison in table I.
The results of Ref. \cite{SHEV} using the AGS formalism are given in
column two and the ones obtained with the formalism of the previous 
section are given in column three. As one can see from table I there
is very good agreement between the two calculations. The small 
discrepancies shown in some cases are of no significance since they
occur for the values of $a_{\eta N}$ allowing the quasibound state
to occur and thereby the solutions
are highly unstable. In our case,  in this situation  we had to use a large 
number of mesh points in order to guarantee stability.

In the multiple-scattering approach of Ref. \cite{GRE1} an approximate
formula was used which is based in a partial sumation of the 
multiple-scattering series. The function $\tau_2(E)$ was taken
to be constant 

\begin{equation}
\tau_2(E)=\lambda_\eta.
\label{eq21}
\end{equation}
Using the relation (\ref{eq19}) this gives

\begin{equation}
\lambda_\eta=-{\alpha_2^4 \over \pi\mu_2}a_{\eta N},
\label{eq22}
\end{equation}
which will be referred to as their model I. They used also a second
model which will be referred to as model II for which instead of Eq.
(\ref{eq22}) they took

\begin{equation}
\lambda_\eta=-{\alpha_2^4 \over \pi\mu_2}{a_{\eta N} \over 1-iq_0 a_{\eta N}},
\label{eq23}
\end{equation}
with $q_0$ = i0.367 fm$^{-1}$. For the nucleon-nucleon interaction
they used a Yamaguchi separable potential with a range parameter
$\alpha_1=1.41$ fm$^-1$.

We compare in table II the results of our exact calculations which 
we obtained using the parameters of Ref. \cite{GRE1} with their 
results using an approximate formula for the two models I and II.
As it can be seen from this table, the approximate formula of
the multiple scattering series works
very well for small values of $a_{\eta N}$, as expected from convergence
arguments. When $a_{\eta N}$ is large the multiple scattering series formula
deviates more from the exact result, nevertheless,
it is still qualitatively 
correct, since it predicts correctly the
quasibound states in all the cases where they exist for both models.

\subsection{$A_{\eta d}$ with separable-potential models}

In the previous subsection we have seen that the models of Refs. 
\cite{SHEV,GRE1} predict a quasibound state if the real part of $a_{\eta N}$
is of the order of 0.7-0.8 fm. However, since their $\eta N$ T-matrix is 
not derived from a potential their function $\tau_2(E)$ is not 
constrained by their form factor $g_2(p_2)$. We will therefore 
construct separable potential models of the coupled $\eta N$ - $\pi N$ 
system that reproduce in one case just the complex scattering length 
$a_{\eta N}$ for arbitrary values of the $\eta N$ 
range-parameter $\alpha_2$ and in another case the full 
$\eta N$-$\eta N$ scattering amplitude around the $S_{11}$ resonance.

Similarly, we will consider two different models of the $NN$ interaction;
a simple Yamaguchi potential that does not have short-range repulsion and
a PEST model which has the same half-off-shell behavior as the Paris
potential so that it contains short-range repulsion. 

If we use in the Lippmann-Schwinger equation of the coupled $\eta N$ -
$\pi N$ system the separable model

\begin{equation}
<p|V_{\eta\eta}|p'>=\lambda_{\eta}g_2(p)g_2(p'),
\label{eq24}
\end{equation}

\begin{equation}
<p|V_{\pi\pi}|p'>=\lambda_{\pi}g_\pi(p)g_\pi(p'),
\label{eq26}
\end{equation}

\begin{equation}
<p|V_{\eta\pi}|p'>=\pm \sqrt{\lambda_{\eta}\lambda_{\pi}}g_2(p)g_\pi(p'),
\label{eq25}
\end{equation}
the T-matrices are of the form

\begin{equation}
<p|t_{\eta\eta}(E)|p'>=g_2(p)\tau_2(E)g_2(p'),
\label{eq28}
\end{equation}

\begin{equation}
<p|t_{\pi\pi}(E)|p'>={\lambda_{\pi}\over \lambda_{\eta}}
g_\pi(p)\tau_2(E)g_\pi(p'),
\label{eq30}
\end{equation}

\begin{equation}
<p|t_{\eta\pi}(E)|p'>=\pm\sqrt{\lambda_{\pi}\over \lambda_{\eta}}
g_2(p)\tau_2(E)g_\pi(p'),
\label{eq29}
\end{equation}
with

\begin{equation}
{1\over \tau_2(E)}={1\over \lambda_{\eta}}-G_2(E)-{\lambda_{\pi}\over
\lambda_{\eta}}G_\pi(E),
\label{eq31}
\end{equation}

\begin{equation}
G_2(E)=\int_0^\infty p^2 dp{g_2^2(p) \over E-p^2/2\mu_2+i\epsilon},
\label{eq32}
\end{equation}

\begin{equation}
G_\pi(E)=\int_0^\infty p^2 dp{g_\pi^2(p) \over E +p_0^2/2\mu_\pi
-p^2/2\mu_\pi+i\epsilon}.
\label{eq33}
\end{equation}
$\mu_2$ and $\mu_\pi$ are the $\eta N$ and $\pi N$ reduced masses 
respectively while $p_0$ is the $\pi N$ relative momentum at the
$\eta N$ threshold, i.e., 

\begin{equation}
p_0^2={[s_0-(m_\pi+m_N)^2][s_0-(m_\pi-m_N)^2] \over 4s_0},
\label{eq34}
\end{equation}
with

\begin{equation}
s_0=(m_\eta+m_N)^2.
\label{eq35}
\end{equation}
If we use simple Yamaguchi form factors

\begin{equation}
g_2(p)={1 \over \alpha_2^2 + p^2},
\label{eq36}
\end{equation}

\begin{equation}
g_\pi(p)={1 \over \alpha_\pi^2 + p^2},
\label{eq37}
\end{equation}
we find that the strengths $\lambda_{\eta}$ and $\lambda_{\pi}$
can be obtained in terms of the real and imaginary parts of
$a_{\eta N}$ as

\begin{equation}
\lambda_{\eta}^{-1}=-{\pi\mu_2\over 2\alpha_2^3}-{\alpha_\pi^2-p_0^2
\over 2\alpha_\pi p_0}{\pi\mu_2\over \alpha_2^4}{Im\,a_{\eta N}\over
|a_{\eta N}|^2}-{\pi\mu_2\over \alpha_2^4}{Re\,a_{\eta N}\over |a_{\eta N}|^2},
\label{eq41}
\end{equation}

\begin{equation}
\lambda_{\pi}=\lambda_{\eta}{\mu_2\over \mu_\pi}
{(\alpha_\pi^2+p_0^2)^2\over p_0\alpha_2^4}{Im\,a_{\eta N}\over
|a_{\eta N}|^2}.
\label{eq42}
\end{equation}
Since we do not include the pion channel explicitly but only through the 
function $\tau_2(E)$ (see Eq. (\ref{eq31})), we will fix the range of the 
$\pi N$ potential to the value $\alpha_\pi=p_0$, for which case the second
term in the r.h.s. of Eq. (\ref{eq41}) drops out and it becomes clear that
in this case the strength of
the $\eta N$ potential is determined by the real part of the $\eta N$ 
scattering length and the strength of the $\pi N$ potential is determined
by the imaginary part of the $\eta N$ scattering length for a given
value of the range $\alpha_2$. Since this models are based in a Yamaguchi
form factor for the $\eta N$ potential we will refer to them as $Y_{\eta N}$
models.

We constructed also more realistic separable models that reproduce not
only the $\eta N$ scattering length but also the most important
features of the $S_{11}$ resonance such as its position and width.
For this we considered the $S_{11}$ amplitudes obtained from the
analyses of Refs. \cite{BAT2,GRE2,GRE3}. We found that with a simple
Yamaguchi model of the $\eta N$ form factor is not possible to
generate a resonance in the $\eta N$ $S_{11}$ channel. We therefore
changed the form factor $g_2(p)$ instead of Eq. (\ref{eq36}) to

\begin{equation}
g_2(p)={A+p^2\over (\alpha_2^2+p^2)^2},
\label{eq46p}
\end{equation}
while keeping for the $\pi N$ form factor the Yamaguchi form (\ref{eq37}).
We give in table III the parameters $\alpha_\pi$, $\lambda_\pi$, $\alpha_2$,
$A$, $\lambda_{\eta}$
of the coupled $\eta N$-$\pi N$ separable
potentials fitted to the $S_{11}$ amplitudes of \cite{BAT2,GRE2} as
well as to the models A, B, C, and D of \cite{GRE3}. 
We show in Fig. 1, as an example, the $\eta N$-$\eta N$ amplitude of
Ref. \cite{BAT2} (dashed lines) compared
with the ones of our separable-potential model. Similar results are
obtained for the other models. Since these models generate a resonance 
in the $\eta N$ channel we will refer to them as $R_{\eta N}$ models.
Since we do not include the pion channel explicitly, only the $t_{\eta \eta}$
component of the coupled $\eta N$-$\pi N$ T-matrix given by
Eq. (\ref{eq28}) has been used after identifying $t_{\eta \eta}$ with
$t_2$ of Eq. (1).

In the case of the $NN$ interaction we have considered two models; the
simple Yamaguchi model used in Ref. \cite{GRE1} which has a range 
parameter $\alpha_1=1.41$ fm$^{-1}$ (which we will refer to as the $Y_{NN}$
model) and the PEST potential constructed
in Ref. \cite{PEST} (which we will refer to as the $P_{NN}$ model) that 
is of the form

\begin{equation}
g_1(p_1)=\sum_{n=1}^6 {C_n\over p_1^2+\beta_n},
\label{eq46}
\end{equation}
where the parameters $C_n$ and $\beta_n$ are given in Ref. \cite{PEST}.
The half-off-shell T-matrix of this separable potential has the same 
behavior as that of the Paris potential and therefore it takes into
account the repulsion at short distances that is present in the
nucleon-nucleon force. 

We give in table IV the results of our separable-potential models for
the $\eta d$ scattering length $A_{\eta d}$ where we have considered all
four combinations of the $\eta N$ and $NN$ separable-potential models.
In the case of the $\eta N$ Yamaguchi model $Y_{\eta N}$ we took the
range parameter $\alpha_2=3.316$ fm$^{-1}$ which is the same as in
Refs. \cite{SHEV,GRE1}. Here however the corresponding T-matrix is
calculated through the Lippmann-Schwinger equation.
In the first column we give the reference for
the $\eta N$ $S_{11}$ amplitude that we used to fit the $R_{\eta N}$ model 
and in the second column we give the $\eta N$ scattering length of that
amplitude which has been used to construct the $Y_{\eta N}$ model. The 
third column gives $A_{\eta d}$ using  simple Yamaguchi models for 
the $\eta N$ and $NN$ interactions and it shows that even with these simple
separable models the quasibound state only appears when $Re\,a_{\eta N}$ 
is about 1.05-1.07 fm while in the multiple-scattering approaches of the
previous section it appeared already with $Re\, a_{\eta N}\approx$ 0.6-0.9 
fm. The fourth column gives the results of the Yamaguchi model for the
$\eta N$ amplitude and the PEST model for the $NN$ amplitude and it
shows that the $NN$ short-range repulsion also works against quasibinding
since it wipes out the quasibound state although the $Im\,A_{\eta d}$ 
remains large. The fifth and sixth columns contain the results of the
resonant model of the $\eta N$ amplitude with Yamaguchi and PEST models
for the $NN$ interaction respectively, and they show that the attraction
of the system is greatly reduced when one takes into account the resonant 
nature of the $\eta N$ amplitude. Notice, however, that experimentally
the effects will be similar whether there is a quasibound state or not 
since in both cases there will be an enhancement of the cross section 
at threshold. 

\subsection{$\eta d$ scattering}

We calculated the integrated elastic cross section of $\eta d$ 
scattering in the region of the $S_{11}$ resonance for the six
resonant models of the $\eta N$ interaction given in table III and the
realistic PEST potential for the $NN$ interaction. We show in Fig. 2 
the results of the three-body model (solid lines) and of the 
impulse approximation (dashed lines). At threshold, the results of all the 
three-body models are about one order of magnitude larger than those 
of the impulse approximation while at higher energies they are 2 or 3 
times smaller. The behavior of the cross section at threshold indicates
that even though the quasibound state is not present, the 
interaction in this region is very strong since it enhances the 
impulse approximation result by about one order of magnitude. Thus, a
signal of this behavior may appear also in other processes  where
there is an $\eta N$ final state like the $np\to \eta d$
reaction where a large enhancement in the cross section has been
observed in the region near threshold \cite{CALE}.

In order to illustrate the effect of the strong $\eta d$ interaction in
the reaction $np \to \eta d$ we will estimate the enhancement of the
$np\to \eta d$ cross section at threshold due to the $\eta d$ rescattering.
We write the amplitude of the process $np\to\eta d$ as

\begin{equation}
A=B+BG_0T_{\eta d},
\label{eq47}
\end{equation}
where $B$ is the amplitude of the $np\to\eta d$ process without $\eta d$
rescattering, $G_0$ is the two-body Lippmann-Schwinger propagator of the 
intermediate $\eta d$ state, and $T_{\eta d}$ is the half-off-shell 
T-matrix of the
elastic $\eta d$ process. If we introduce the $\eta d$ elastic-scattering 
amplitude
$F_{\eta d}=-\pi\mu_{\eta d}T_{\eta d}$, where $\mu_{\eta d}$ is the 
$\eta d$ reduced mass,
then at threshold, Eq. (\ref{eq47}) is written explicitly as

\begin{equation}
A=B(0)[1+{2\over\pi}\int_0^\infty dq_1{B(q_1)\over B(0)}F_{\eta d}(q_1)],
\label{eq48}
\end{equation}
where $F_{\eta d}(q_1)$ is given by Eq. (\ref{eq10}) with $q_{10}$ 
replaced by $q_1$. Therefore, the enhancement factor of the $np\to\eta d$
cross section due to $\eta d$ rescattering is

\begin{equation}
f=|1+{2\over\pi}\int_0^\infty dq_1{B(q_1)\over B(0)}F_{\eta d}(q_1)|^2.
\label{eq49}
\end{equation}

The amplitude $B$ of the $np\to\eta d$ process without $\eta d$ rescattering
is presumably given by meson exchanges such as $\pi$, $\rho$, and $\eta$
followed by the excitation and decay of the $S_{11}$ resonance \cite{WILK}. The
explicit form of the production operator without $\eta d$ rescattering is
not so important since in our estimate of the enhancement factor given by 
Eq. (\ref{eq49}) only the ratio $B(q_1)/B(0)$ enters. Therefore, we 
take for it the $\eta$-exchange amplitude generated by our three-body model,
i.e., 

\begin{equation}
B(q_1)=\int_0^\infty q_2^2 dq_2\, K_{12}(q_1,q_2;E)\tau_2(E-q_2^2/2\nu_2)
K_{23}(q_2,q_3;E),
\label{eq50}
\end{equation}
where $E=-B_d$ and $q_3=\sqrt{(m_\eta+m_d)^2-m_N^2}$ is the momentum 
corresponding to an initial $NN$ state. Using the models 1-6 of Fig. 2,
taken from references
\cite{BAT2,GRE2,GRE3} and included in Table IV,
we obtained the enhancement factors $f$ = 2.5, 2.7, 3.1, 3.3, 4.7, and
5.1 respectively. These estimates are quite comparable to the enhancement factors
observed in Ref. \cite{CALE}, in special for the 4 first models.

\section{Conclusions}

Recently, the  experimental band for the $\eta N$ scattering
length $a_{\eta N}$ has been pushed towards larger values for its
real part \cite{GRE2,GRE3}. For those larger values, the $\eta N$ models
not generated
directly from an integral equation, which would fix naturally their off-energy-shell
behavior needed in three-body calculations, predict a quasi-bound $\eta
NN$ state. If instead, the new data is used to generate $\eta$N t-matrices
calculated from a potential and an integral equation, our results indicate
that a realistic NN interaction, like the Paris potential, through its
short-range correlations, prevents the existence
of the bound-state, independently of the $\eta$N models, provided they
have been built dynamically. We confirmed then that 
the predictions of a $\eta NN$ eta-mesic nucleus are crucially affected by the
off-shell behavior of the underlying $\eta$-N models.
Importantly, however, is that even for an inexisting quasi-bound state,
an exact three-body calculation for the multi-scattering series in the final
state predicts a severe enhancement of the elastic $\eta d$ cross-section
in the narrow region from threshold to 5-10 MeV above threshold.
This result is independent of the $\eta N$ two-body
models. Very likely, the enhancement predicted by the exact three body 
calculations
is related to the one observed
in the reactions $np \to \eta d$ \cite{CALE} and $\gamma d \to \eta d$ 
\cite{METAG,FIX}. We actually made an estimate of the enhancement of the
cross section of $np \to \eta d$ due to the $\eta d$ final state interaction.
Within the three-body model of our work,
this enhancement is in the ball park of the empirical findings of \cite{CALE}.

\acknowledgements
This work was supported in part by 
COFAA-IPN (M\'exico) and
by Funda\c c\~ao para a Ci\^encia e a Tecnologia,
MCT (Portugal) under contracts PRAXIS XXI/BCC/18975/98 and
PRAXIS/P/FIS/10031/1998.

\begin{figure}
\caption{ Real and imaginary parts of the $\eta N\to \eta N$ $S_{11}$
amplitude. The solid and dashed lines are the results of our resonant
$R_{\eta N}$ separable potential model 1; the symbols give the amplitudes
of Ref. [10].}
\end{figure}

\begin{figure}
\caption{Integrated elastic cross sections of the three-body model
(solid lines) and of the impulse approximation (dashed lines)
for the six resonant $\eta N$ models of table III using for the
$NN$ interaction the PEST model, as a function of the c.m. $\eta d$ kinetic
energy.}
\end{figure}

\begin{table}
\caption{$\eta d$ scattering length $A_{\eta d}$ (in fm)
as calculated in Ref. [4] with the AGS formalism and as calculated here
with the Faddeev formalism for different values
of the $\eta N$ scattering length $a_{\eta N}$ (in fm).}

\begin{tabular}{ccccc}
 & $a_{\eta N}$   & AGS  [4] & Faddeev & \\
\tableline
 & 0.25+i0.16  & 0.73+i0.56  & 0.73+i0.56  & \\
 & 0.30+i0.30  & 0.61+i1.22  & 0.61+i1.22  & \\
 & 0.46+i0.29  & 1.31+i1.99  & 1.33+i1.99  & \\
 & 0.579+i0.399  & 0.34+i3.31  & 0.34+i3.33  & \\ 
 & 0.876+i0.274  & -8.81+i4.30  & -9.02+i4.15  & \\
 & 0.98+i0.37  & -4.69+i1.59  & -4.74+i1.53  & \\
\end{tabular}
\end{table}

\begin{table}
\caption{$\eta d$ scattering length (in fm) for models I 
and II calculated in
Ref. [3] using a multiple-scattering
theory (MST) and as calculated here using the Faddeev formalism
for different values
of the $\eta N$ scattering length $a_{\eta N}$ (in fm).}

\begin{tabular}{ccccccc}
& $a_{\eta N}$  & MST I [3] & Faddeev 
& MST II [3]  & Faddeev  &  \\
\tableline
 & 0.25+i0.16  & 0.66+i0.71  & 0.64+i0.71  & 0.66+i0.58  & 0.65+i0.58  & \\
 & 0.30+i0.30  & 0.39+i1.28  & 0.38+i1.25  & 0.58+i1.11  & 0.56+i1.09  & \\
 & 0.46+i0.29  & 0.72+i2.04  & 0.62+i1.95  & 1.11+i1.54  & 1.04+i1.54  & \\
 & 0.579+i0.399  & -0.13+i2.64 & -0.08+i2.31  & 0.93+i2.41 & 0.74+i2.28 &\\
 & 0.876+i0.274  & -2.76+i4.24 & -1.54+i2.55  & 2.42+i5.55 & 0.46+i5.21 &\\
 & 0.98+i0.37 & -2.75+i2.77  & -1.16+i2.05 & -0.06+i6.20  & -1.12+i4.21 &\\
\end{tabular}
\end{table}

\begin{table}
\caption{Parameters of the $\eta N$-$\pi N$ separable potential 
models fitted to the $S_{11}$ resonant amplitudes given in
Refs. [10-12].}

\begin{tabular}{cccccccccc}
& Model & Ref. & $a_{\eta N}$& $\alpha_{\pi}$   & $\lambda_{\pi}$  & $\alpha_2$  
& $A$  & $\lambda_\eta$ &  \\
\tableline
& 1 &[6] & 0.72+i0.26 &   1.2       
  &   -0.028013
  &   15.5    
  &   21.394872     
  &   -8817.102932 & \\
& 2 &[7]  & 0.75 + i0.27 &  1.2    
  &    -0.028013  
  &    14.5 
  &    18.657140 
  &    -7222.670614 &\\
& 3 &[8](D)  &  0.83 + i0.27 &  1.2     
  &    -0.028013    
  &    13.68 
  &    17.7754 
  &    -5970.573595  &\\
& 4 &[8](A)  &  0.87+i0.27 &  1.2  
  &    -0.028013 
  &    13.1 
  &    16.646837 
  &    -5215.563508  & \\
& 5 &[8](B)  & 1.05 + i0.27 &  1.2      
  &    -0.028013     
  &    11.1 
  &    13.357879  
  &    -3080.397824  &\\
& 6 &[8](C)  & 1.07 + i0.26 &   1.2     
  &    -0.028013  
  &    10.47 
  &    11.952401 
  &    -2579.494411  &\\
\end{tabular}
\end{table}

\begin{table}
\caption{$\eta d$ scattering length (in fm) for various models of the 
$\eta N$ and $NN$ interactions where the respective t-matrices were
calculated through a dynamical equation. $Y_{\eta N}$ stands for the
Yamaguchi model of the $\eta N$ interaction with a range parameter
$\alpha_2=3.316$ fm$^{-1}$, $R_{\eta N}$ stands
for the resonant model of the $\eta N$ interaction, $Y_{NN}$ stands
for the Yamaguchi model of the $NN$ interaction, and $P_{NN}$ stands
for the PEST model of the $NN$ interaction. The first column indicates
the reference on which the resonant $\eta N$ model is based and the
second column indicates the $\eta N$ scattering length (in fm) of that model.}  

\begin{tabular}{cccccccc}
& Ref. & $a_{\eta N}$ &  $Y_{\eta N}Y_{NN}$  & $Y_{\eta N}P_{NN}$  
& $R_{\eta N}Y_{NN}$  & $R_{\eta N}P_{NN}$ &  \\
\tableline
&[6]    & 0.72+i0.26 & 2.38+i3.04 & 2.59+i2.37 & 2.43+i1.60 & 2.46+i1.62& \\
&[7]    & 0.75+i0.27 & 2.38+i3.41 & 2.70+i2.64 & 2.56+i1.65 & 2.61+i1.72& \\
&[8](D) & 0.83+i0.27 & 2.57+i4.51 & 3.21+i3.34 & 3.00+i1.87 & 3.10+i2.03& \\
&[8](A) & 0.87+i0.27 & 2.54+i5.20 & 3.47+i3.78 & 3.20+i1.94 & 3.36+i2.19& \\
&[8](B) & 1.05+i0.27 & 0.06+i8.66 & 4.27+i7.05 & 4.20+i2.24 & 4.81+i3.19& \\
&[8](C) & 1.07+i0.26 &-0.42+i9.21 & 4.53+i7.60 & 4.21+i2.04 & 5.02+i3.14& \\
\end{tabular}
\end{table}

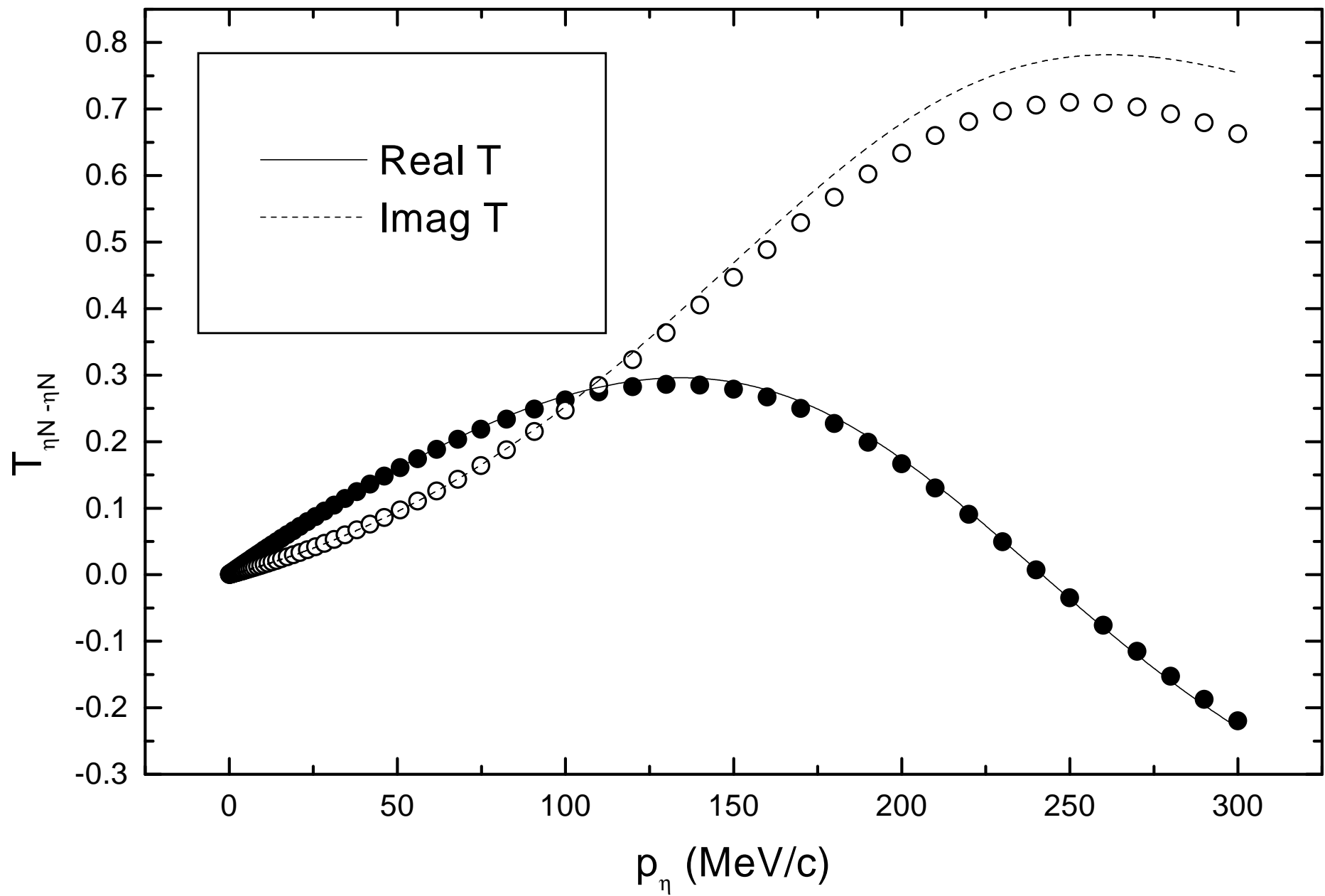
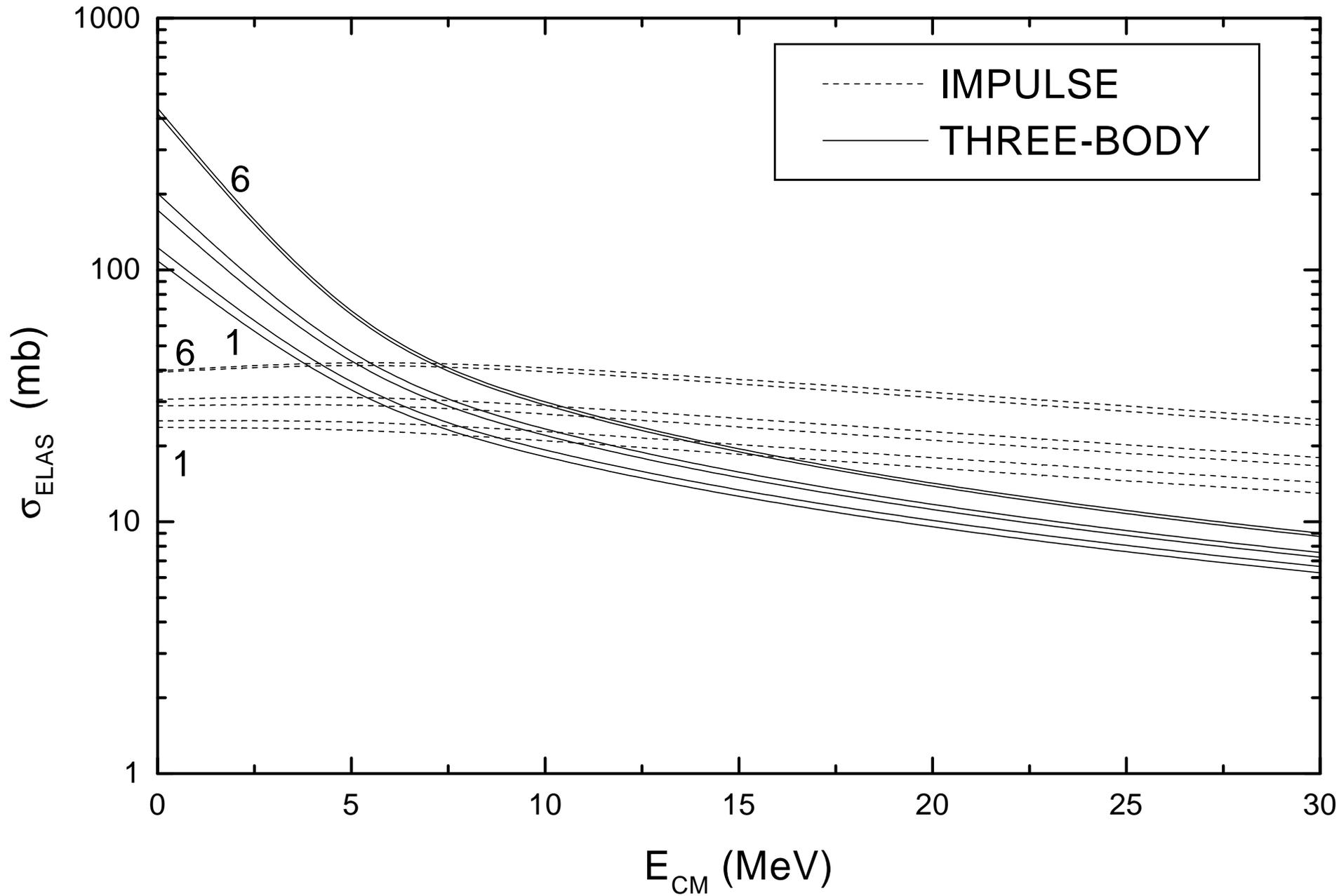
\end{document}